\documentclass[conference,a4paper]{IEEEtran}
\IEEEoverridecommandlockouts

\makeatletter
\def\markboth#1#2{\def\leftmark{\@IEEEcompsoconly{\sffamily}\MakeUppercase{\protect#1}}%
\def\rightmark{\@IEEEcompsoconly{\sffamily}\MakeUppercase{\protect#2}}}
\makeatother

\usepackage[english]{babel}
\usepackage{xcolor}
\usepackage{lipsum}% http://ctan.org/pkg/lipsum
\usepackage{caption}
\usepackage{cite}
\usepackage[pdftex]{graphicx}
\usepackage{subcaption}
\usepackage{amsmath}
\usepackage{booktabs}
\usepackage{colortbl}
\usepackage{amsfonts}
\usepackage{amssymb}
\usepackage{amsthm}
\usepackage{array}
\usepackage{verbatim}
\usepackage{listings}
\usepackage{algorithm}
\usepackage{algpseudocode}
\usepackage{url}
\usepackage{enumerate}
\usepackage{multirow}

\usepackage{epsfig}
\usepackage{epstopdf}
\usepackage{multicol}% http://ctan.org/pkg/multicols
\usepackage[font=footnotesize]{caption}

\newcommand{\bi}{\begin{itemize}}
\newcommand{\ei}{\end{itemize}}
\newcommand{\be}{\begin{equation}}
\newcommand{\ee}{\end{equation}}

\def\beq{\begin{equation}}
\def\eeq{\end{equation}}
\def\beqa{\begin{eqnarray}}
\def\eeqa{\end{eqnarray}}
\def\beqan{\begin{eqnarray*}}
\def\eeqan{\end{eqnarray*}}

\title{The Bufferbloat Problem  \\ over Intermittent Multi-Gbps mmWave Links}
\author{{{\bf Menglei Zhang}$^\diamond$, {\bf Marco Mezzavilla}$^\diamond$, {\bf Jing Zhu}$^\dagger$, {\textbf{Sundeep Rangan}$^\diamond$}, {\bf Shivendra Panwar}$^\diamond$ }\\
$^\diamond$ NYU WIRELESS, Brooklyn, NY, USA \qquad\qquad
$^\dagger$ Intel, Santa Clara, USA\\
emails: \small{$\{$\texttt{menglei}, \texttt{mezzavilla}, \texttt{srangan}, \texttt{panwar}$\}$\texttt{@nyu.edu}, \texttt{jing.z.zhu@intel.com}
}}

\begin{document}
\maketitle

\begin{abstract}

Due to massive available spectrum in the millimeter wave (mmWave) bands,
cellular systems in these frequencies
may provides orders of magnitude greater capacity than networks in
conventional lower frequency bands.
However, due to high susceptibility to blocking, mmWave
links can be extremely intermittent in quality.
This combination of high peak throughputs and intermittency can cause
significant challenges in end-to-end transport-layer mechanisms such as
TCP.  This paper studies the particularly challenging problem of
bufferbloat.  Specifically, with current buffering and congestion
control mechanisms, high throughput-high variable links can
lead to excessive buffers incurring long latency.
In this paper, we capture the performance trends obtained while adopting two potential solutions that have been proposed in the literature: Active queue management (AQM) and dynamic receive window. We show that, over mmWave links, AQM mitigates the latency but cannot deliver high throughput. The main reason relies on the fact that the current congestion control was not designed to cope with high data rates with sudden change. Conversely, the dynamic receive window approach is more responsive and therefore supports higher channel utilization while mitigating the delay, thus representing a viable solution.
\end{abstract}
\smallskip
\begin{IEEEkeywords}
5G, millimeter wave communication, cellular systems, AQM, congestion control
\end{IEEEkeywords}

%\section{Introduction}
%\label{introduction}

\section{Introduction}
\label{bufferbloat}

The millimeter wave (mmWave) bands -- roughly corresponding to
frequencies above 10~GHz --  have attracted considerable attention for
next-generation cellular wireless systems
~\cite{KhanPi:11-CommMag,rappaportmillimeter,RanRapE:14,andrews2014will,ghosh2014millimeter}.
The bands offer orders of magnitude more spectrum than conventional cellular
frequencies below 3 GHz
-- up to 200 times by some estimates \cite{KhanPi:11-CommMag}.
The massive bandwidth can be
combined with the large number of spatial degrees of freedom available in
high-dimensional antenna arrays to enable cellular systems
with orders of magnitude greater capacity \cite{AkdenizCapacity:14,BaiHeath:14,nix1}.

At the same time, mmWave links are likely to have highly variable quality.
MmWave signals are completely blocked by many common
building materials such as brick and mortar,
\cite{Allen:94,Anderson04,Alejos:08,singh2007millimeter,KhanPi:11-CommMag,Rappaport:28NYCPenetratioNLOSs},
and even the human body can cause
up to 35~dB of attenuation \cite{LuSCP:12}.
As a result, the movement of obstacles and reflectors,
or even changes in the orientation of a handset relative to the body or a hand,
can cause the channel to rapidly appear or disappear.

As a consequence, mmWave signals have the unique feature of having
extremely high peak rates combined with high variability.
This combination is extremely challenging when viewed from an end-to-end perspective
\cite{zhang2016transport}.  Specifically, transport layer mechanisms
and buffering must rapidly adapt to the link capacities that can dramatically
change.  This work addresses one particularly important problem -- bufferbloat.

\paragraph*{Bufferbloat}

Bufferbloat is triggered by persistently filled or full buffers, and usually results in long latency and packet drops. This phenomenon was first pointed out in late 2010\cite{gettys2011bufferbloat}. Optimal buffer sizes should equal the bandwidth delay product (BDP), however, as the delay is usually hard to estimate, larger buffers are deployed to prevent losses.
Even though these oversized buffer prevent packet loss, the overall performance degrades, especially  when transmitting TCP flows,\footnote{TCP carries almost 90$\%$ of the internet traffic\cite{lan2006measurement}.} which is the main focus of this paper. Originally, TCP was designed to react and adjust its sending rate based on timely congestion notifications, e. g., as a function of the packet drop rate. However, the oversized buffer concealed the congestion from TCP, resulting in high sending window values, which determine the maximum packets that can be send out without acknowledgments (ACKs), also called \emph{packets-in-flight}. The problem of oversized buffers begins when the sending window grows beyond capacity, thus generating buffering delays. In our previous work\cite{zhang2016transport}, we showed that sending TCP packets over intermittent and high peak
capacity mmWave links resulted in (i) severe latency trends with large buffers, and (ii) low throughput due to TCP retransmissions with small buffers. In this paper, we address the bufferbloat problem by investigating and evaluating two existing solutions.

\paragraph*{Challenges for mmWave}

It is well known that the mmWave channel usually has large bandwidth, and can support very high (Multi-Gbps) data rate, especially with Line-of-Sight (LoS). On the other hand, due to e2e congestion control, the throughput of a TCP connection is limited by TCP send window size as well as round trip time.  In cellular systems, the round trip time (RTT)
may be large due to the need to route through the core network to a
packet gateway.
Hence, in order to fully utilize the mmWave channel, it is critical for the TCP transmitter to maintain a very large TCP send window. For example, let us assume data rate = 3Gbps and RTT = 40ms, leading to the BDP of 15MB. As a result, TCP send window must always stay above 15MB in order to achieve the maximum e2e throughput. However, when packet loss happens due to congestion or any other reason, TCP sender will trigger congestion avoidance and reduce its send window by half. Afterwards, it takes one RTT for the send window to increase by 1 segment. If one TCP segment is 1K bytes long, it will take 40 seconds to increase TCP send window by 1MB! For example, if the TCP send window is 10MB when congestion happens, it will take 200 seconds for the TCP send window to increase from 10MB to 15MB, large enough to achieve the maximize e2e throughput. All in all, it is challenging to fully utilize the Multi-Gbps mmWave channel with TCP traffic.

The problem is made particularly important due to the variability of the channel.
As mentioned above, mmWave links can rapidly change in quality.
The TCP window will thus need to rapidly increase or decrease to track the channel
fluctuations and maintain an appropriate window size.  Otherwise, buffers can
either bloat or have a queue underflow.

%The first challenge is high propagation loss, results limited signal transmitting range. Feasible solutions might includes shrinking cell size and beaforming technique. Another challenge is the vulnerability to blockages. From our measurements, the receiving power may suddenly drop around 15 dB or 30 dB when blocked by humans or buildings, respectively. This distinct property makes the communication through mmWave link extremely unstable -- many packets might be corrupted when transmitting via the deteriorated wireless link. Even though the lost packets can be recovered by link layer retransmissions and large buffer at base station, bufferbloat issue becomes more serve as mentioned in \cite{zhang2016transport}. Even though attain ultra low latency while maintaining high utilization over intermittent high capacity channel seems to be quit challenging, some solutions are proposed to be able to mitigate the bufferbloat while maintaining same throughput.
%\subsection{Semi-empirical propagation channel model}
%We introduced NYU channel model in [paper], which models the blockage events as sudden increase of propagation loss. To better understand the actual channel dynamics, we use sounding equipment to continuously measure the receiving power while people are walking by. After collecting realistic blockage measurement traces, we developed a new semi-empirical channel model which superimposes the NYU channel model and realistic traces, and enabled us to simulate small scare fading, along with smooth LoS-NLoS-LoS scenario transitioning.

\paragraph*{End-to-end mmWave simulation}
In order to fully capture the mmWave challenges described above, we harness the end-to-end simulation framework \cite{mmW-ns3} based on the network simulator ns-3 \cite{ns3}. This module, which has been developed internally, includes: 
\begin{itemize}
\item A detailed characterization of the mmWave channel, which can be generated through statistical models or real traces obtained with our channel sounder;
\item Antenna array model and beamforming capabilities;
\item A flexible and customizable frame structure at the MAC layer;
\item 4G LTE standard-compliant functionalities from the RLC layer above (including the evolved packet core (EPC) \cite{LENA-lte}).
\end{itemize}
Thanks to this framework \cite{mswim, e2e-ns3-mmw}, we have shared the first TCP performance trends over mmWave-aided cellular networks in \cite{tcp-menglei}, which motivated the writing of this paper.

\paragraph*{Contributions}
The two main contributions of this paper are (i) the first of its kind performance evaluation of AQM techniques over mmWave and (ii) the implementation of a novel cross-layer algorithm that successfully mitigates the bufferbloat problem while delivering high throughput. 

\section{Active Queue Management}
\label{aqm}

Typical queue management techniques involve single queue, first in first out (FIFO) and Drop-tail. Even though Drop-tail is easy to manage, it may cause unnecessary delay: As the queue is building up, the round trip time (RTT) also increases.

AQM is a promising solution to address the bufferbloat issue in wireless networks. It reacts to congestion much faster, by dropping packets when operating at certain regimes, to mitigate the increased latency effect. Some early AQM, such as random early detection (RED) \cite{floyd1993random}, were widely studied in the literature, but failed to find market traction because of the intrinsic complexity of its tuning parameters. Recently, a simpler AQM technique, namely CoDel~\cite{nichols2012controlling}, was proposed to replace RED queues, and adapt to dynamic link rates without parameter configuration. However, there are no contributions exploring the AQM performance in 5G mmWave cellular system, which is one of the goals of our paper. CoDel is able to discriminate ``good" and ``bad" queues: good queues can quickly empty the buffer, whereas ``bad" queues do persistently buffer packets. It works by monitoring the minimum queue delay in every 100 ms interval,\footnote{This is a default parameter that can be changed. The author claimed that these parameters are optimal over any link} and only drop packets when the minimum queue delay is more than 5 ms.

We compare the performance of Drop-tail and CoDel queues in two scenarios, where a mobile UE is experiencing blockages from (i) other humans or (ii) buildings. The main difference is that, with humans, the channel deteriorates slowly and the blockage lasts a short interval; on the other hand, with buildings, the link capacity drops rapidly and the blocking interval is much longer. These trends are captured in Fig. \ref{fig:CoDelH} and Fig. \ref{fig:CoDelB}, respectively. The sender opens a FTP connection and sends a large file to the UE. The congestion control is TCP Cubic, with delayed ACK disabled. The maximum queue length is 50k packets. The core network latency is 40 ms.

\subsection{Human blockage}

\begin{figure}[t!]
    \centering
    \includegraphics[trim={2.5cm 1.4cm 3.5cm 0cm},clip, width=0.5\textwidth]{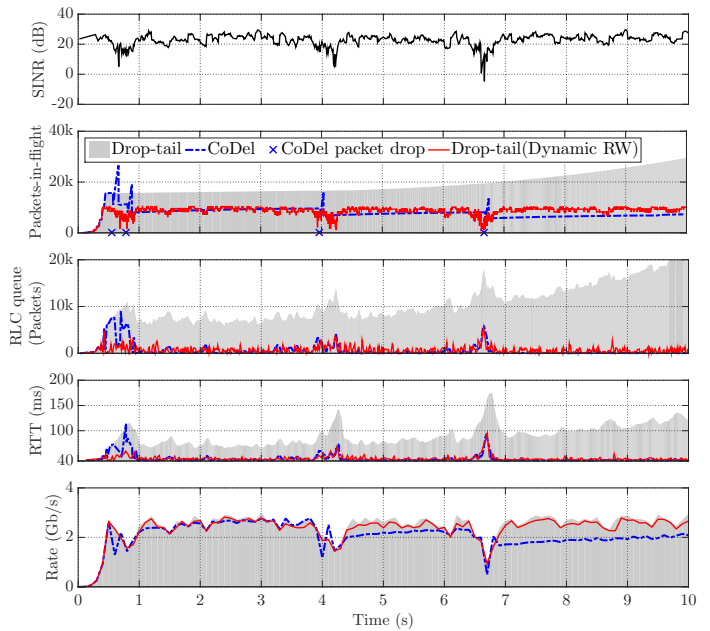}
    \caption{Single UE with human blockages}
    \label{fig:CoDelH}
\end{figure}

In this scenario, the UE is walking at 1 m/s, 300 meters away from the base station, while maintaining LoS connectivity, and experiencing 3 human blocking events. The blockage events were simulated by superimposing the real human blockage measurement traces measured with our sounding equipments \cite{giordani}, over the channel models obtained in \cite{AkdenizCapacity:14}.

\textbf{Drop-tail:} Since the RLC queue size is large enough and all packets lost in the wireless link are recovered by means of lower layer retransmissions (RLC ARQ and MAC HARQ), the sender is unaware of the packet loss, thus keeping a large congestion window that results in high throughput, but also high buffer occupancy and consequent high delay.

\textbf{CoDel:} CoDel has the ability to actively drop packets when it detects high buffering delay. The CoDel packet drop events are also labeled in Fig.\ref{fig:CoDelH}. At 0.5 s, as the RLC queue is building up, the first packet is dropped, which informs the sender to reduce the congestion window. At 0.8 s, the human blockage deteriorates the wireless link capacity and causes the RLC queue to grow thus triggering one more packet drop. Similarly, at 4 s and 6.7 s, two more packets are dropped. The consequent congestion window decrease cleared out all the RLC buffered packets. Nonetheless, when the wireless link recovered from human blockage, the congestion window ramps up to link capacity very slowly.

\subsection{Building blockage}
In this scenario, the UE is walking at 1 m/s, 150 meters away from the base station. The link transitions from LoS to NLoS when crossing a building. The blockage duration is roughly 2 s. The time-characterization of the channel quality drop loss during the transition between LoS and NLoS was obtained from our measurements, as reported in \cite{giordani}.

\textbf{Drop-tail:} As manifested in the previous case, there is no packet drop event observed at the sender. This entails (i) large congestion window, (ii) high buffer occupancy, and (iii) large delay. It is interesting to note how, due to the sudden capacity drop, the RLC buffer grows dramatically resulting in very high latency values.

\begin{figure}[t!]
    \centering
    \includegraphics[trim={2.5cm 1.4cm 3.5cm 0cm},clip, width=0.5\textwidth]{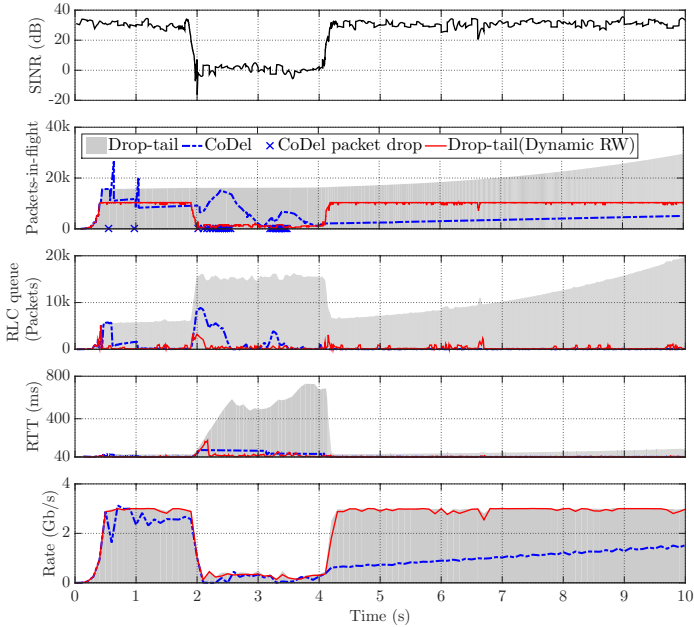}
    \caption{Single UE with building obstacle}
    \label{fig:CoDelB}
\end{figure}

\textbf{CoDel:} In this scenario, CoDel inefficiency is even more severe. During the transition to NLoS, multiple packets were dropped to force the sender backing off and mitigate the fast growing queue. This resulted in near zero throughput because the fast retransmission takes too long to recover multiple drops. This active queue technique, as observed in the previous scenario, dramatically affect the TCP ramp up time after the blockage event, as shown in Fig. \ref{fig:CoDelB}.

\section{Dynamic Receive Window}
\label{drw}

\begin{figure*}[t!]
    \centering
    \includegraphics[trim={3cm 2.5cm 3cm 2cm},clip,width=0.68\textwidth]{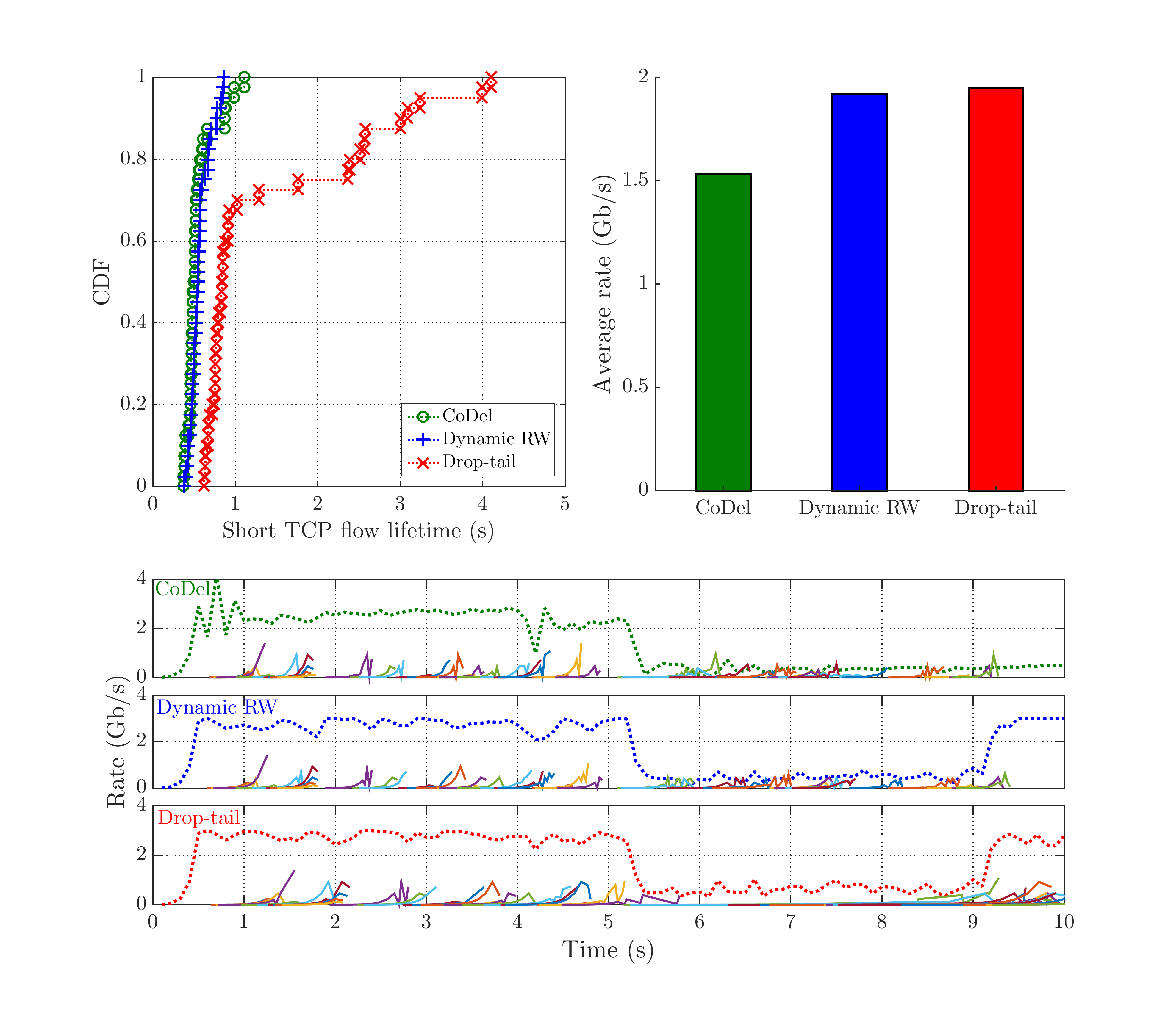}
    \caption{Application rate of long flow (dashed curve) and short flows}
    \label{fig:flow}
\end{figure*}

Before implementing any congestion control, senders used to inject packets into the network as demanded. These unregulated flows seriously damaged the network performance, given that packets would be buffered at the bottleneck link router, resulting in large queueing delays and buffer saturation.

The TCP protocol was introduced to solve this issue by letting the sender slowly probing the available bandwidth and regulating the sending rate. The amount of data delivered by the sender is equal to $min(CW,RW)$. $CW$ is the congestion window, i.e., the amount of bytes/packets in flight without ACKs, determined at the transmitter side - which is based on the TCP variant. $RW$, instead, represents the receive window, i.e., the available receive buffer size piggybacked to the sender. Nowadays, the receive buffer size becomes relatively large and is almost never limiting the sending rate. A recent work \cite{jiang2012understanding} shows that some mobile devices, instead of sending back the available buffer size, they select different RW values based on the connected network, e.g., if it connects to Wi-Fi, large RW size is used, but as it handovers to cellular, smaller RW values are selected.

The authors in \cite{jiang2012tackling} and \cite{liumitigating} showed that informing the sender with the optimal RW, substantially reduces latency without deteriorating the throughput. Further, as all the changes are made at the receiver side, this approach can be easily deployed.

With dynamic receive window Adjustment (DRWA) \cite{jiang2012tackling}, the RW is only based on the RTT and does not exploit channel information. On the other hand, available bandwidth based receiver window dynamic adjustment (ABRWDA) \cite{liumitigating} encapsulates the wireless link capacity while feeding information back to the sender. However, because some wireless resources are reserved for broadcasting, control messages, pilots, etc., the actual wireless link capacity overestimates the available data rate. Even though selecting larger RW values never reduces the utilization in current networks, multi-Gbps pipes introduced at mmWave bands will suffer from large delay.
Further, even if the receiver may be able to extract the precise capacity for data, this value is still overestimated when the channel is shared by multiple UEs. Hence, we propose a new mechanism to better estimate the available capacity and consequently perform a better RW estimation, which equals the optimal bandwidth-delay product.

\textbf{Optimal bandwidth:} Thanks to downlink control messages (DCI) messages, which contain the transport block (TB) size -- the effective number of bits that will be delivered to each UE, users can estimate the allocated bandwidth. Like noted above, if we use the entire bandwidth, such as ABRWDA, it overestimates the RW when multiple UEs are active. On the other hand, if we feed back the effective allocated bandwidth, if the UE capacity suddenly drops, the sender limits its delivery rate thus underutilizing the wireless link when it transition to a better condition. When the congestion did not take place in the wireless link, the base station also allocates less resources to the UE, and the UE feeds back a smaller RW. When the congestion is gone, the sender is still limited by the small RW, and entails low utilization. Therefore, picking either one as the reference bandwidth is not optimal. We propose to use the entire bandwidth when the RTT is within a low latency region, which is $[RTTmin, RTTmin + \delta]$,\footnote{In our simulations, we set $\delta$ to be 10 ms because it showed good performance in terms of utilization and delay.} since the TCP socket can infer there is no bufferbloat issue. Conversely, if the RTT is not operating in the low latency region, the allocated bandwidth is selected to have a more conservative sending rate, in order to mitigate the delay.

\textbf{Optimal delay:} Our approach applies the same method to measure the receiver side RTT when TCP timestamp is on as mentioned in \cite{jiang2012tackling}. In order to prevent over-inflating the RW, we should avoid using the end to end latency to compute the RW. The correct latency should be the delay between the remote host to the UE with an empty buffer. Similar as \cite{liumitigating}, one simple solution is selecting the min RTT. It is reasonable to assume that, if no multi-path TCP is used, the core network latency should be relatively stable and by selecting the min RTT, we are able to find the RTT of of an empty buffer.

We conducted the same experiments performed in Sec. \ref{aqm}, as reported in Fig.\ref{fig:CoDelH} and Fig.\ref{fig:CoDelB}. The RW is dynamically updated based on the optimal bandwidth-delay product. The DRW outperforms CoDel by having much higher throughput and roughly the same delay.
%The only issue we observed is when the channel suddenly deteriorates due to blockage,  in Fig.\ref{fig:CoDelH}. The reaction is delayed by one RTT and results in a delay jitter in the RTT plot shows in Fig. \ref{fig:CoDelH}.

\section{Additional scenarios}

\subsection{Short flows} A realistic scenario would be a user web browsing and texting while a background file is downloading. To test how dynamic RW mechanism improves the user experience, we simulated a long TCP flow along with some short TCP flows, which are randomly distributed. We repeated this experiment with normal RW (both Drop-tail and CoDel queue) and dynamic RW. The rate plot is given in Fig. \ref{fig:flow}. The result shows that the both dynamic RW and CoDel might be able to reduce the delay, but only dynamic RW still maintaining high throughput.

\subsection{Multiple Users}
\begin{figure}[t!]
    \centering
    \includegraphics[trim={2.5cm 0.5cm 3cm 1cm},clip, width=0.45\textwidth]{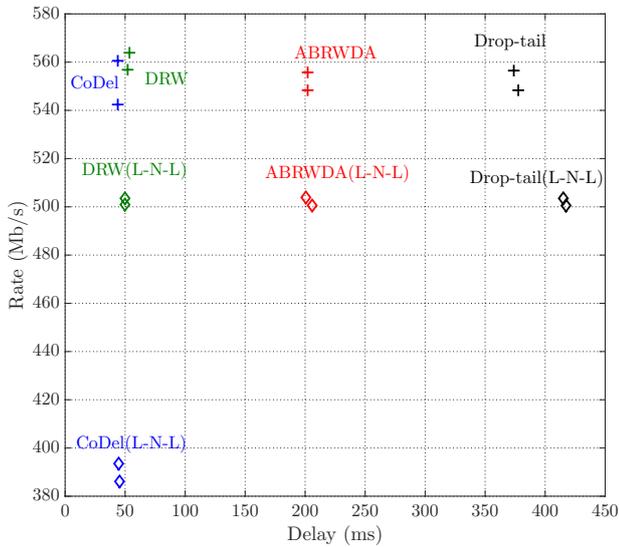}
    \caption{2 LoS UE, 2 LoS-NLoS-LoS UE}
    \label{fig:ratevsdelay}
\end{figure}

In the previous section we showed that DRW outperforms Drop-tail and CoDel in single UE case. Note that in single UE scenario, ABRWDA should have similar performance as DRW since allocated bandwidth almost equals total data bandwidth. In this section, we study the behavior of multiple UEs  connected to the same base station. We established 4 connections between 4 remote hosts and 4 UEs, the TCP flows go through the same base station. 2 UEs are always LoS, 2 UEs experience LoS-NLoS-LoS transitions. The resources are allocated to UEs with Round-robin scheduling decision. The average throughput versus delay is plotted in Fig. \ref{fig:ratevsdelay}: For the 2 LoS UEs, CoDel and DRW have the least delay and almost the same throughput compared to the other two methods. Nonetheless, for the 2 LoS-NLoS-LoS UEs, the DRW shows the best performance.

\begin{figure}[t!]
    \centering
    \includegraphics[trim={2.5cm 0.5cm 3cm 1cm},clip, width=0.45\textwidth]{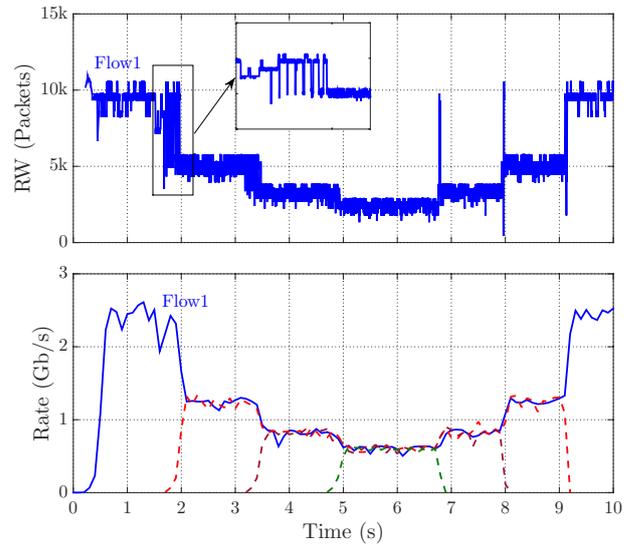}
    \caption{Multiple UE connection}
    \label{fig:arrival}
\end{figure}

To discover the reason why DRW can achieve such a low latency while maintaining good throughput, we conducted the following experiment. In one cell, 1 UE is always connecting, 3 other UEs joined and left the cell at different moments. Each UE established one TCP flow and Fig. \ref{fig:arrival} shows how the RW of flow 1 reacts when other UEs join or leave the cell. When a new UE arrives, the RW starts bouncing between the upper bound (RTTmin times total data bandwidth) and lower bound (RTTmin times allocated bandwidth) and finally becomes stable as the rate of new UE ramps up and share more bandwidth. At 7 s, 8 s, and 9 s, when a UE leaves, the delay quickly reduces and the RW jumps back to the upper bound. Due to the RW inflating behavior, all remaining UEs informed the sender to inject more packets and cause the base station increase to the allocated bandwidth for all the remaining UEs. Because the RW is larger than the optimal window now, the delay will also increase and cause the RW falls back to the new lower bound -- bandwidth divided by the remaining UEs, as shown Fig. \ref{fig:arrival}.

\subsection{Bufferbloat over Uplink}

In the uplink, the UE is the sender.  As a result, the link capacity 
is known at the UE side itself (via the DCI allocations).  
Hence, if a cross-layer design is possible, where the UE MAC layer 
information can be exposed to the TCP sender on the same device, the
TCP sender can directly adjust the congestion window.  
Simulations of this mechanism is a possible future avenue of research.

%TDMA is proposed to be the potential multiple access of mmWave cellular. Due to the symmetric design, the bufferbloat might happen in UL when the users are uploading large files. It can be mitigated by applying the same AQM mechanism as in the DL case.

\section{Conclusions}

We have presented the first comprehensive evaluation of bufferbloat
on an end-to-end simulation of mmWave cellular links.  Our simulation methodology employs
realistic, detailed measurement-based channel models.  Importantly, these models
can capture dynamics in the channel due to the blockage and transitions from LoS to NLoS
states, which is the main problem in end-to-end performance.  The evaluation
also has complete models of the MAC, RLC and networking layers.  

Our study finds that bufferbloat can be severe problem
for mmWave cellular systems due to the high variability of the channel
combined with the delays in the cellular core network.  Moreover,
conventional AQM techniques are unable to mitigate the bufferbloat problems.  
In contrast, we find dynamic receive window can greatly reduce the delay
with minimal loss in throughput.  The main challenge is to enable some form
of cross-layer design. Specifically, the proposed algorithm requires exposing
MAC layer information (DL or UL grants in the DCI messages) to the TCP process
at the UE. How best to do this is one possible avenue of future work.
Nevertheless, our findings present a promising initial result that
properly using channel information at the UE can dramatically improve end-to-end
performance with relatively simple changes.

\label{conclusion}

\bibliographystyle{IEEEtran}
\bibliography{biblio}

\end{document}